\documentclass[review,3p,onecolumn,number,11pt]{elsarticle}

\usepackage{lineno}
\usepackage{multirow}
\usepackage[normalem]{ulem}

\biboptions{sort&compress,square}

\usepackage{lineno}
\usepackage[colorlinks=true,allcolors=blue]{hyperref}
\modulolinenumbers[5]
\usepackage[latin1]{inputenc}
\usepackage{amsmath,empheq}
\usepackage[most]{tcolorbox}
\usepackage{amsfonts}
\usepackage{amssymb}
\usepackage{subfig}
\usepackage{graphicx}
\usepackage[font=small,format=plain,labelfont=bf,textfont=normal,justification=justified,singlelinecheck=false,tableposition=top]{caption}
\usepackage{natbib}
\usepackage{float}
\usepackage{longtable}
\usepackage{color}
\usepackage{booktabs}
\usepackage{makecell}
\usepackage{siunitx}

\newcommand{\figref}[1]{Fig.~\ref{#1}}
\newcommand{\eqnref}[1]{Eqn.~\ref{#1}}
\newcommand{\sectref}[1]{\S~\ref{#1}}

\usepackage[textsize=tiny,backgroundcolor=green!40]{todonotes}

\usepackage{enumitem}
\setlist[description]{leftmargin=1cm,labelindent=1cm}

\DeclareMathOperator{\Div}{\nabla \cdot}
\DeclareMathOperator{\Grad}{\nabla}

\newcommand{\cref}{c_{\mathrm{ref}}}

\journal{--}

\begin{document}

\begin{frontmatter}

\title{Shape matters: Understanding the effect of electrode geometry on cell resistance and chemo-mechanical stress}



\author[mymainaddress]{Tiras Y. Lin\corref{mycorrespondingauthor}}
\ead{lin46@llnl.gov}
\author[mymainaddress]{Hanyu Li}
\author[mymainaddress]{Nicholas W. Brady}
\author[mymainaddress]{Nicholas R. Cross}
\author[mymainaddress]{Victoria M. Ehlinger}
\author[mymainaddress]{Thomas Roy}
\author[mymainaddress]{Daniel Tortorelli}
\author[mymainaddress]{Christine Orme}
\author[mymainaddress]{Marcus A. Worsley}
\author[mymainaddress]{Giovanna Bucci\corref{mycorrespondingauthor}}
\cortext[mycorrespondingauthor]{Corresponding author}
\ead{bucci3@llnl.gov}

\address[mymainaddress]{Lawrence Livermore National Laboratory,
7000 East Ave, Livermore, CA 94550, USA}

\begin{abstract}
Rechargeable batteries that incorporate shaped three-dimensional electrodes have been shown to have increased power and energy densities for a given footprint area when compared to a conventional geometry, i.e., a planar cathode and anode that sandwich an electrolyte. Electrodes can be shaped to enable a higher loading of active material, while keeping the ion transport distance small, however, the relationship between electrical and mechanical performance remains poorly understood. A variety of electrode shapes have been explored, where the electrodes are individually shaped or intertwined with one another. Advances in manufacturing and shape and topology optimization have made such designs a reality. In this paper, we explore sinusoidal half cells and interdigitated full cells. First, we use a simple electrostatics model to understand the cell resistance as a function of shape. We focus on low-temperature conditions, where the electrolyte conductivity decreases and the governing dimensionless parameters change. Next, we use a chemo-mechanics model to examine the stress concentrations that arise due to intercalation-driven volume expansion. We show that shaped electrodes provide a significant reduction in resistance, however, they result in unfavorable stress concentrations. Overall, we find that the fully interdigitated electrodes may provide the best balance with respect to this trade-off.
\end{abstract}




\end{frontmatter}

\section{Introduction}
Improving the performance of electrochemical energy storage devices has the potential to impact a bevy of applications, such as wearable electronics, electric vehicles, and grid-scale energy storage \cite{gur2018review}. 
As a result, significant work has focused on improving material characteristics, such as the electrolyte conductivity and electrode microstructure composition.
Another electrode property, which is of particular interest here, is the macroscale shape of the battery.
``Shaping''  battery electrodes yields improvements in both power and energy densities as compared to conventional planar electrodes \cite{long2004three}. 
What is relatively less understood, however, is how these shaped batteries perform in more extreme conditions. 
For example, in space flight and applications involving harsh weather, it is critical to understand how these electrodes perform under thermal and mechanical loads, and in low-temperature environments, where electrolyte conductivities drop significantly \cite{landesfeind2019temperature,hwang2018ionic,valoen2005transport,feng2017evaluating}. 
To increase technology readiness levels (TRL), it is crucial to understand the effect of electrode shape on the cell resistance, in particular at low temperature, and on the mechanical stress that develops from electrode expansion due to either intercalation or thermal load.

Three-dimensional shaped electrodes have been shown to be able to optimize the inherent trade-off between capacity and rate performance present in conventional planar electrodes by reducing the distance of transport pathways while increasing the active material loading \cite{long20203d}.
This is particularly beneficial in scenarios where ionic conductivity in the electrolyte is limiting, e.g. in liquid electrolytes at low temperature, solid electrolytes at room temperature, thick electrodes, and electrodes with significant tortuosity. 
A variety of electrode shapes have been proposed and categorized into different classes by \citet{hung2022three}: electrodes can be (1) either porous or non-porous; and (2) either individually shaped and separated by a planar separator or completely intertwined or interpenetrating with one another. 
The interdigitated design, where protruding cylindrical rods or planar fins from each electrode sandwich the electrolyte to allow for a thin electrolyte gap, has emerged as a popular design \cite{hung2021modeling,zadin2011finite,zadin2010modelling}. 
Other geometries, such as aperiodic ``sponge'' and concentric cylinder geometries, where the electrolyte coats one shaped electrode with the other electrode filling in the remaining gaps \cite{long2004three,long2003ultrathin,rhodes2004nanoscale}, and electrodes with periodic microcolumns \cite{mckelvey2017microscale}, have also been explored. 
A common trait of these complex geometries, however, is that they have highly nonuniform current distributions \cite{hart20033}, which may lead to poor electrode utilization.

The growth of shape and topology optimization (TO), coupled with rapid advances in additive manufacturing, has increased the opportunity to explore shapes that are more intricate and complex, such as the hierarchical teeth-like structures observed by \citet{roy2022topology} for porous electrodes and the tree-root structures observed by \citet{lin2022topology} for flow manifolds.
TO distributes material in a design domain to minimize some specified cost function \cite{sigmund2013topology}, such as power losses or structure compliance. 
TO has previously been used to design flow systems \cite{dilgen2018density,de2021three,gersborg2005topology,borrvall2003topology} and various structural systems \cite{bendsoe1989optimal,bruns2001topology}.
These optimization approaches automate the design process, avoiding tedious trial-and-error, wherein users manually iterate on designs using expensive simulations. 
However, optimization is a relatively nascent  approach for designing electrochemical systems where many physical phenomena are coupled together. 
That said, it has been used to design flow fields for flow batteries \cite{lin2022topology,yaji2018topology,chen2019computational} and fuel cells \cite{behrou2019topology,wang2023enhancing,qu2023design}, supercapacitors \cite{batista2023design}, and nonuniform porosity redox electrodes \cite{beck2021computational}. 
Notably, \citet{roy2022topology} and \citet{li2024topology} have used TO to design porous electrodes for half- and full-cell energy storage devices, respectively. 
While some optimized designs exhibit intuitive shapes, unexpected design features also appear, due to the vast design space afforded by TO. 
A physics-based understanding of battery shapes is necessary to justify these complex designs.

Although shaped electrodes continue to improve the electrical performance of batteries, it is unclear how these designs perform mechanically, particularly for solid-state electrolytes, which, in comparison to a liquid electrolyte, does not deform freely. 
As Li ions intercalate and deintercalate, the electrodes expand and contract, leading to stress concentrations. 
While there have been prior studies on the chemo-mechanical stress arising in intercalating electrodes at the particle-level \cite{zhang2008intercalation,zhang2007numerical} and composite-level \cite{xu2019heterogeneous}, less work has been done to predict the stress at the macroscale level, i.e., over the entire shaped electrodes. 
Physically, this behavior is similar to that of confined thermal expansion, where materials expand and contract due to temperature changes \cite{hetnarski2009thermal,eslami2013theory}. 
Prior work has modeled the stress arising within cooled multilayer elastic media \cite{hsueh1985residual, hsueh2002thermal} and structural beams. 
However, these studies have not been extended to  battery electrode applications.
This is needed to determine whether the electrodes shaped to increase their power or energy density are mechanically viable.

In this paper, we leverage relatively simple models to explore the shape effects of porous electrodes with respect to cell resistance and mechanical stability. 
In \sectref{sec:electrostatics}, we model non-planar half cell and full cell porous electrodes.
We focus on low-temperature operating conditions, where the electrolyte conductivity drops significantly, and use dimensionless groups to illustrate the benefits of shaping under these conditions.
In \sectref{sec:mech}, we explore the corresponding chemo-mechanics of the shaped electrodes, to model the stress concentrations that develop during expansion and contraction. 
We discuss the propensity for the shaped electrodes to fail and the trade-off between their superior electrical and inferior mechanical performance. 
Finally, in \sectref{sec:conc}, we conclude with a discussion on potential consequences for battery design and areas for future research.

\section{Electrostatics of shaped electrodes} \label{sec:electrostatics}
In this section, we consider the effect of electrode shape on the cell overpotential using a simplified electrostatics model. 
We investigate half cell and full cell designs, to understand how non-planar electrode geometries can compensate for the kinetic limitations of materials and interfaces.  
We show that electrode shaping is especially beneficial for low-temperature operating conditions when the transport properties of the electrolyte degrade.
In the half cell, the interface between the porous electrode and electrolyte region is defined by a sinusoidal function (\figref{fi:schematic}(A)), and for the full cell, we explore interdigitated structures (\figref{fi:schematic}(B)).
To ensure a fair comparison, in all cases, the volumes occupied by the porous electrode and the pure electrolyte regions are identical.
We also assume identical microstructures, i.e., active-material particles, binder, carbon additives, and pore-filled electrolyte, and identical electrochemical surface areas (ECSA) per unit volume.

\subsection{Electrostatics model}

\begin{figure*}[t]
\centering
\makebox[\textwidth][c]{\includegraphics[width=7.25in]{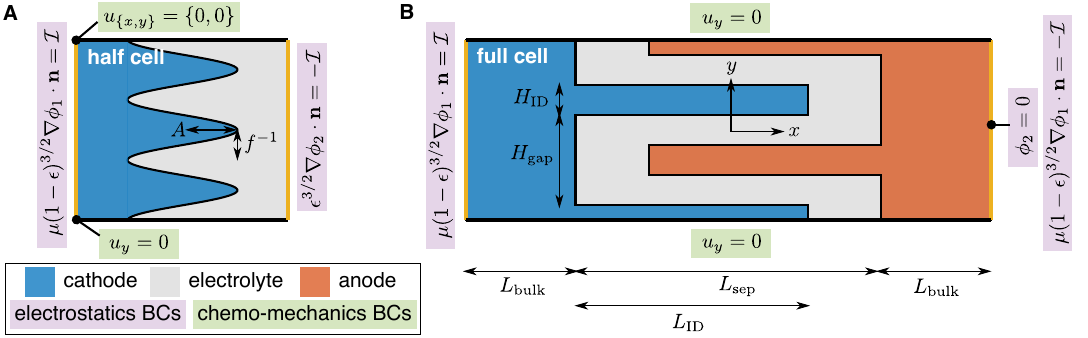}}
\caption{Schematics of the (A) half cell and the (B) full cell. Only nonzero Neumann and Dirichlet boundary conditions corresponding to the electrostatics  and the chemo-mechanics models are shown. The total dimensional domain sizes are $2L\times 2L$ and $4L\times 2L$ for the half and full cells, respectively, where $L=100$ $\mu$m.}
\label{fi:schematic}
\end{figure*}

We consider the electrode/electrolyte system shown in \figref{fi:schematic}. The governing equations for the electrostatic potentials are,
\begin{align}
\Div (\sigma \Grad \phi_1) &= a i_n (\phi_1, \phi_2), \\ 
\Div (\kappa \Grad \phi_2) &= - a i_n (\phi_1, \phi_2),
\end{align}
where $\phi_1$ and $\phi_2$ are the solid-phase and liquid-phase potentials, respectively, $\sigma$ and $\kappa$ are the effective solid-phase and liquid-phase conductivities, $a$ is the specific surface area of the electrode (ECSA per unit volume), and $i_n$ is the current density. In general, we consider boundary conditions corresponding to galvanostatic conditions, representing a constant applied current density $I$ on the current collector with all other boundaries insulated, i.e.
\begin{align}
\sigma \Grad \phi_1 \cdot\mathbf{n} = \pm I,  \quad \Grad \phi_2 \cdot\mathbf{n}=0 & \text{ on current collector},\\
\Grad \phi_1 \cdot\mathbf{n}=\Grad \phi_2 \cdot\mathbf{n}=0 & \text{ everywhere else},
\end{align}
where $\mathbf{n}$ is the outward surface normal vector.

For Butler-Volmer kinetics, the current density is modeled as a function of the local overpotential, $\eta=\phi_1-\phi_2-U_0$,
\begin{equation}
i_n (\phi_1, \phi_2) = i_0\left[ \frac{c_a}{\cref} \exp{\left(\frac{\alpha_a F}{RT} \eta \right)} - \frac{c_c}{\cref}\left(-\frac{\alpha_c F}{RT} \eta \right) \right],
\end{equation}
where $F$ is Faraday's constant, $R$ is the ideal gas constant, and $T$ is the temperature. In this study, we assume that the standard potential $U_0=0$ and the transfer coefficient $\alpha=\alpha_a=\alpha_c=0.5$. Additionally, we neglect mass transfer effects by setting the electrolyte concentration $c=c_a=c_c$ to be a constant. Of course, neglecting mass transfer is not reasonable for all battery conditions. We restrict our attention to the case when either (1) the transference number $t_+=z_+u_+/(z_+u_+-z_-u_+)$ of the reactive ion, e.g. Li$^+$, where $z_i$ and $u_i$ are the charges and mobilities of the ions, is near one \citep{fuller2018electrochemical}, or (2) when the time after closing the battery circuit is short. Both of these assumptions would limit the concentration gradients, and hence mass transfer effects. It is important to note that, since the model used in this work is rather general, our analysis could potentially be applied to other systems, such as flow batteries. To further simplify the analysis, we linearize the kinetics assuming that the local overpotential is small, so that
\begin{equation}
i_n (\phi_1, \phi_2) \approx i_0 \frac{c}{\cref}\frac{F}{RT}(\phi_1-\phi_2),
\end{equation}
which is consistent with the assumption of a small concentration gradient. 

To account for the effect of porosity, we use the Bruggeman correlation \citep{tjaden2016origin}, so that 
\begin{equation} \label{eq:brugg}
\sigma = \sigma_0 (1-\epsilon)^{3/2}, \quad \kappa = \kappa_0 \epsilon^{3/2},
\end{equation}
where $\epsilon$ is the porosity, and $\sigma_0$ and $\kappa_0$ are the pure solid and liquid conductivities, respectively. In the porous electrode region, $0<\epsilon<1$, and $\epsilon=1$ in the pure electrolyte region. 

To nondimensionalize, we assign
\begin{equation}
(\hat{x}, \hat{y}) = \left(\frac{x}{L},\frac{y}{L}\right), \quad \hat{\phi}_{1,2}=\phi_{1,2} \frac{F}{RT},
\end{equation}
where $L=100$ $\mu \mathrm{m}$ is the assumed thickness of a half cell. By doing so, we obtain the dimensionless groups
\begin{align} 
\mu=\frac{\sigma_0}{\kappa_0},& \quad \mathrm{Wa}=\frac{\kappa_0 RT}{i_0 L F}, \label{eq:dimgroups1} \\
\rho= a L,&\quad \mathcal{I}=\frac{L}{\kappa_0}\frac{F}{RT}I,\quad \mathcal{C}=\frac{c}{\cref}, \label{eq:dimgroups2}
\end{align}
where $\mu$ is the ratio of solid to liquid conductivity, $\mathrm{Wa}$ is the Wagner number, $\rho$ is a measure of the roughness of the porous electrode (i.e., the ratio between the electrochemical surface area and the geometric surface area),  $\mathcal{I}$ is the dimensionless applied current, and $\mathcal{C}$ is the dimensionless concentration. 
As discussed previously, we neglect mass transfer, and thus we assume a spatially uniform value of $\mathcal{C} = 1$.

After nondimensionalizing and removing the notation $(\hat{\cdot})$ for simplicity, the governing equations with linearized kinetics are
\begin{align}
\Div \left(\mu \left(1-\epsilon \right)^{3/2} \Grad \phi_1 \right) &= \frac{\mathcal{C} \rho}{\mathrm{Wa}} (\phi_1-\phi_2), \label{eq:phi1_nondim} \\ 
\Div \left(\epsilon^{3/2} \Grad \phi_2 \right) &= - \frac{\mathcal{C} \rho}{\mathrm{Wa}} (\phi_1-\phi_2), \label{eq:phi2_nondim}
\end{align}
with galvanostatic boundary conditions, which are summarized in \figref{fi:schematic}. Boundary conditions that are not explicitly specified are zero Neumann conditions. The problem considered here closely follows the work of \citet{newman1962theoretical}, who considered a planar electrode. 
This model thus predicts the concentration-independent secondary current distribution for a shaped porous electrode.

We focus on the effects of varying the Wagner number $\mathrm{Wa}$ and  conductivity ratio $\mu$. 
Therefore, we first discuss  the physical underpinnings of these groups. 
The Wagner number represents the ratio between the surface resistance $R_s$ and Ohmic resistance $R_\Omega$, i.e. $\mathrm{Wa}=R_s/R_\Omega$, where
\begin{equation}
    R_s  = \frac{\partial \eta}{\partial i_n} = \frac{1}{i_0} \frac{RT}{F} \frac{\cref}{c},\qquad
    R_{\Omega} = \frac{L}{\kappa_0}. 
\end{equation}
The Wagner number is an indicator for the uniformity of current distribution:  a larger $\mathrm{Wa}$ corresponds to a more uniform current distribution throughout the porous electrode. 
Note that in our simplified model in \eqnref{eq:phi1_nondim}-\eqnref{eq:phi2_nondim}, $\mathrm{Wa}$ is divided by the surface roughness $\rho$ and concentration $\mathcal{C}$. 
Thus, a larger electrochemical surface area, and hence, larger $\rho$, reduces the effective $\mathrm{Wa}$, i.e. $\mathrm{Wa}/(\mathcal{C}\rho)$. 
Additionally, while we do not account for mass transfer effects here, we see that larger concentrations $\mathcal{C}$ also reduce the effective $\mathrm{Wa}$. 
Similarly, we see that strong gradients in $\rho$ and $\mathcal{C}$ would make this effective Wagner number spatially dependent.

The conductivity ratio $\mu$ defined in \eqnref{eq:dimgroups1} controls where the majority of the current distribution occurs: for small $\mu$, the current predominantly occurs at the current collector-porous electrode interface, and for large $\mu$, the current predominantly occurs at the porous electrode-bulk electrolyte interface \citep{newman2021electrochemical}.
In other words, for $\mu \ll 1$, the low electronic conductivity of the solid phase limits the reaction kinetics such that the charge-transfer reaction occurs in the vicinity of the current collector. 
Conversely, for $\mu \gg 1$, ion transport in the electrolyte limits the kinetics such that the charge-transfer reaction front occurs in the vicinity of the electrode-electrolyte interface.

\begin{figure*}[t]
    \centering
    \makebox[\textwidth][c]{\includegraphics[width=7in]{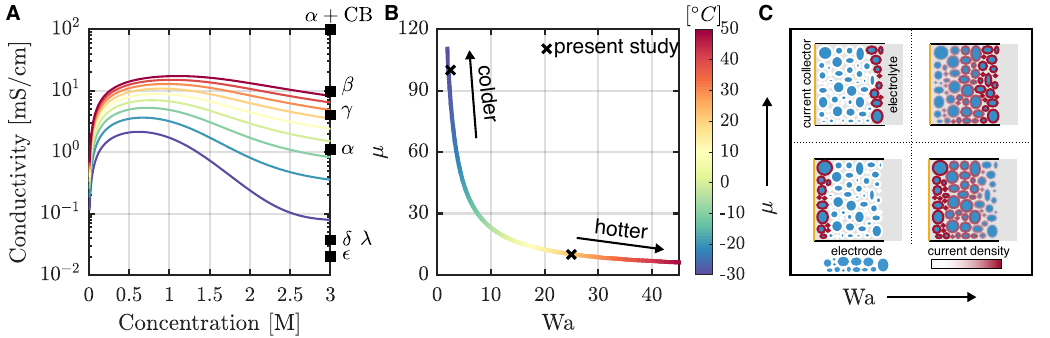}}
    \caption{(A) Electrolyte conductivity correlation (curves) from \citet{landesfeind2019temperature} and conductivity of MnO$_2$ phases (squares) from \citet{xiao2023ultrahigh}, where an order-of-magnitude estimate of the conductivity when carbon black (Super-P) is added \citep{cho2015comparative} is also shown. (B) $\mu$ vs $\mathrm{Wa}$ as a function of temperature. (C) A qualitative schematic of how the current distribution in the porous electrode changes with $(\mu,\mathrm{Wa})$. }
    \label{fig:params}
\end{figure*}

We aim to understand how low-temperature battery operation  reduces kinetics, and how electrode shaping can compensate for this reduction.
In \figref{fig:params}(A), we show the electrolyte conductivity correlation (curves) described by \citet{landesfeind2019temperature}, for LiPF$_6$ EC:DMC (1:1 w:w), as a function of temperature and electrolyte concentration. 
At 1 M concentration, the conductivity decreases by almost one order of magnitude when the temperature drops from room temperature (\qty{20}{\celsius}) to \qty{-30}{\celsius}.
Additionally, \figref{fig:params}(A) shows the conductivity $\sigma_0$ (squares) of various phases of MnO$_2$, as reported by \citet{xiao2023ultrahigh}. 
Our study assumes that the electric conductivity of the solid electrodes is temperature independent in the $-30$ to \qty{50}{\celsius} range, since the dependence is significantly weaker \cite{buerschaper1944thermal}. In comparison to other cathode materials, the electric conductivity of MnO$_2$ is low,  and as such, conductive additives are typically added to the MnO$_2$ electrode. Small amounts of carbon additives are effective in improving the electric conductivity of composite electrodes. As an example,  \figref{fig:params}(A)   shows an order-of-magnitude estimate of the conductivity when carbon black (Super-P) is added to $\alpha-$MnO$_2$ \citep{cho2015comparative}, indicated by the label $\alpha+\text{CB}$.
Our choices of an LiPF$_6$ EC:DMC electrolyte material and an MnO$_2$ cathode material are based on their popularity and on the fact that their combined properties allow us to explore a wide range of (Wa, $\mu$) values. Nonetheless, our results have general validity and are applicable to other battery chemistries.

To begin our study, we first estimate the magnitudes of $\mathrm{Wa}$ and $\mu$. The exchange current $i_0$ is estimated to be $i_0=\qty{10}{A/m^2}$ based on the expression provided by \citet{girard2015enhancing}. 
In reality, it is possible that the exchange current $i_0$ may depend on temperature, and this would affect the resulting dimensionless groups. In this study, however,  we assume $i_0$ is a constant, due to a lack of data to support a specific form of temperature dependence. Note however, that since the electrostatics analysis in this work is dimensionless, our results could in principle be reinterpreted to account for this. 
Next, the specific surface area is estimated as \cite{weng2018modeling}
\begin{equation}
a=3(1-\epsilon)/r_p,
\end{equation}
where the particles that are assumed to comprise the electrode have a radius $r_p=1.5$ $\mu$m. Using these estimates, and the temperature-dependent liquid conductivities from \figref{fig:params}(A), in \figref{fig:params}(B), we show the curve that is traced out between $(\mu, \mathrm{Wa})$ as the temperature is varied based on the definitions in \eqnref{eq:dimgroups2}.
We consider room temperature (\qty{20}{\celsius}) and low temperature (\qty{-30}{\celsius}) conditions, whereby $(\mu, \mathrm{Wa})=(10,25)$ and $(\mu, \mathrm{Wa})=(100,2.5)$, respectively. Finally, in \figref{fig:params}(C), we qualitatively show  the current distribution within the porous electrode as ($\mu$, $\mathrm{Wa}$) changes. As $\mu$ increases, the reaction shifts from the current collector to the electrolyte interface, and as $\mathrm{Wa}$ increases, the reaction becomes more uniform with respect to the macroscale features. At high $\mathrm{Wa}$, the current is localized at the individual electrode-particle interfaces, and its distribution is controlled by microscale features.

The cell electrostatics model is used to determine the effect of shape on cell overpotential,
\begin{equation}
\eta_\mathrm{cell} = \langle \phi_1 \rangle_\mathrm{left} - \phi_{2,\mathrm{right}},
\end{equation}
where $\langle \phi_1 \rangle_\mathrm{left}$ is the average solid potential over the current collector of the electrode on the left, and $\phi_{2,\mathrm{right}}=0$ is the liquid potential at the right boundary, i.e. the electrolyte boundary for the half cell, and the current collector of the electrode on the right for the full cell. The cell overpotential $\eta_\mathrm{cell}$ is a proxy for the cell resistance. Note that the specific reference value of $\phi_{2,\mathrm{right}}$ does not matter, as long as it is fixed somewhere in the domain. \eqnref{eq:phi1_nondim}-\eqnref{eq:phi2_nondim} are solved in Wolfram Mathematica using the finite element method with \texttt{NDSolveValue}.

\subsection{Results and discussion}

\subsubsection*{Half-cell analysis}

\begin{figure*}[t]
    \centering
    \makebox[\textwidth][c]{\includegraphics[width=7in]{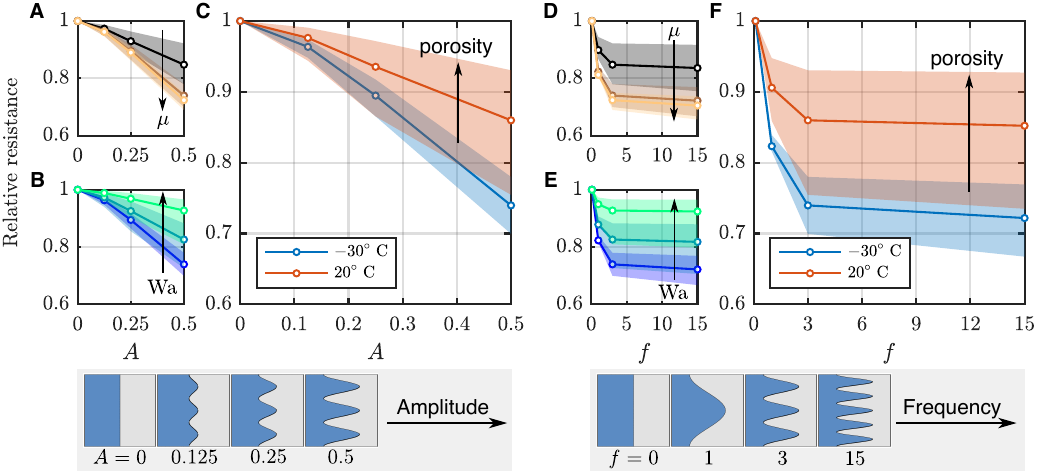}}
    \caption{The relative resistance of the cell as a function of the  (A, B, C) amplitude $A$ and (D, E, F) frequency $f$ of the interface. In panels (A, D), the Wagner number $\mathrm{Wa}$ is fixed at $2.5$, and we vary the solid to liquid conductivity ratio $\mu\in\{10,100,1000\}$, and in panels (B, E), $\mu$ is fixed at $100$, and we vary $\mathrm{Wa}\in\{2.5, 25, 250\}$. In panels (C, F), $\mu$ and $\mathrm{Wa}$ are chosen to correspond to low and room temperature conditions. The colored bands correspond to the range of porosities tested, with $\epsilon=0.5 \pm 0.2$.} 
    \label{fig:resistance_temp_halfcell}
\end{figure*}

For the half cell analyses, cf.  \figref{fi:schematic}(A), the shape of the interface is described by the function 
\begin{equation} 
    x = A \cos \left(f \pi y \right),
    \label{eq:shape_halfCell}
\end{equation}
where $A$ and $f$ are the dimensionless amplitude and frequency, and the total dimensionless domain length is $2$.
First, we explore the effect of varying $A$ for fixed $f=3$ and a variety of conductivity ratios $\mu$ and Wagner numbers $\mathrm{Wa}$. 
In \figref{fig:resistance_temp_halfcell}(A, B), we explore the effect of varying $A$ between $0$ and $0.5$ for fixed $f=3$  on the relative resistance, for a variety of conductivity ratios $\mu$ and Wagner numbers $\mathrm{Wa}$. 
The lines indicate the cell resistance values for a fixed electrode porosity $\epsilon = 0.5$, whereas the shaded regions indicate the variation of resistance for $\epsilon \in [0.3,0.7]$.
For low $\mathrm{Wa}$, the benefits of having larger features increases as $\mu$ increases, although the improvement saturates for $\mu>100$.
Low porosity electrodes receive the most benefit as $A$ increases.
When $\mathrm{Wa}\approx 1$ and $\mu\gg 1$, the majority of the current density accumulates over the electrode/bulk electrolyte interface region, and as a result, ion transport is enhanced with the increasing interfacial area, i.e., increasing $A$. 
On the other hand, when $\mathrm{Wa}\gg 1$,  the results are much less sensitive to $A$, since the current density distribution is more uniform (\figref{fig:resistance_temp_halfcell}(B)). 
In other words, the larger the $\mathrm{Wa}$, the larger the $\mu$ would have to be to see a benefit of shaping. 
This explanation is consistent with the \figref{fi:schematic}(C) discussion.
Note that when $\mathrm{Wa}\approx 1$ and $\mu\ll 1$, it may be of interest to consider shaping the current collector, i.e. where the current density would be high, to enhance electron transport.

Next, we focus specifically on the ($\mu$, $\mathrm{Wa}$) pairs  corresponding to low and room temperature operation. 
In  \figref{fig:resistance_temp_halfcell}(C, F), we show the relative resistance as a function of amplitude $A$ and frequency $f$ for these two cases. 
Consistent with the previous discussion, we see the benefit of shaping, both with respect to amplitude and frequency, to be more significant in the low temperature case, where Wa $\approx 1$ and $\mu\gg 1$.  
Again, electrode shaping is more beneficial at lower porosity, as the effective ionic transport in the porous electrode decreases according to the Bruggeman relationship (\eqnref{eq:brugg}) and the effective electric conductivity increases.
We find that while the resistance continues to decrease with amplitude -- until $A=1$, where presumably, the battery shorts since the electrode spans the entire domain length -- the resistance plateaus as frequency is increased. This is because, as $f\rightarrow\infty$, 
the transport properties of the shaped electrode become equivalent to a planar electrode with an increased porosity of $2\epsilon$.
This variable porosity slab electrode is consistent with experiments that show that graded porosity electrodes, wherein the highest porosity region contacts the electrolyte, have improved mass transport, leading to higher capacity retention with charging rate and cycle number~\cite{yang2022gradient}.
In \figref{fig:resistance_temp_halfcell}(D, E), we also show the resistance as a function of frequency $f$ for a variety of  $\mu$ and $\mathrm{Wa}$, and the results are consistent with the above discussion.

\subsubsection*{Full-cell analysis}
\begin{figure*}
    \centering
    \makebox[\textwidth][c]{\includegraphics[width=7.in]{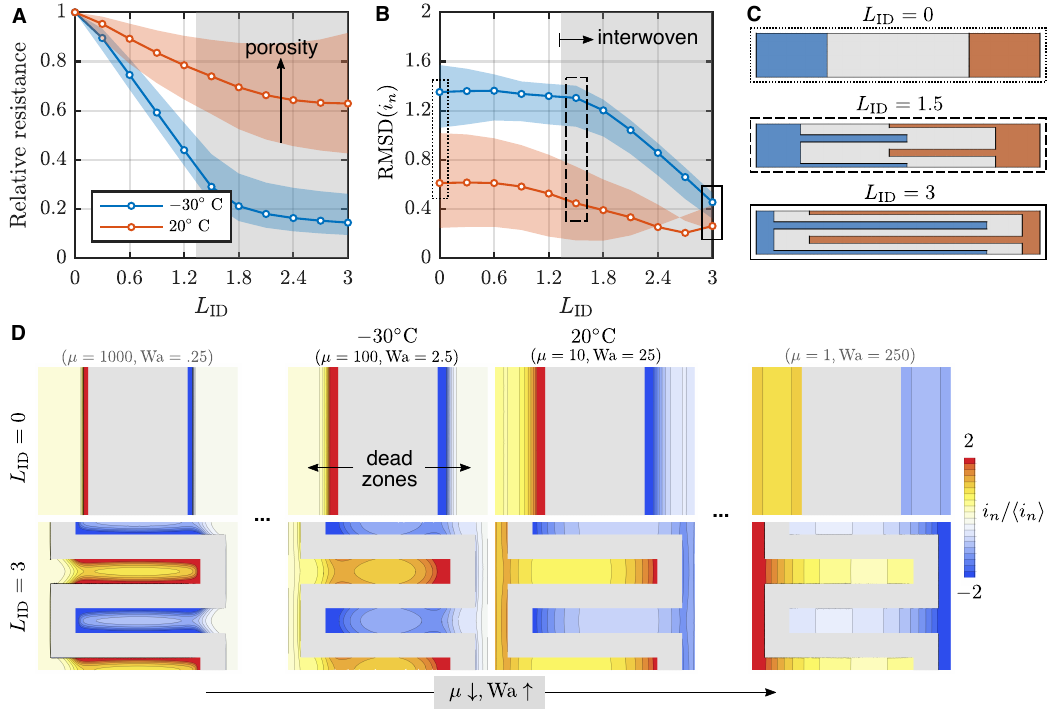}}
    \caption{The (A) relative resistance and (B) root-mean-squared-deviation $\mathrm{RMSD}(i_n)$ of the full cell as a function of the length of the interdigitated fin $L_{\mathrm{ID}}$ for low and room temperature. In panel (B), note that the trend with porosity is not always monotonic in porosity $\epsilon$. (C) A schematic of how the geometry appears as the $L_{\mathrm{ID}}$ increases, where the black boxes correspond to the data in panel (B). The (D) current distribution $i_n/\langle i_n\rangle$ is shown for low and room temperature cases for the full cell with planar electrodes and fully interdigitated electrodes. To illustrate the trend, additional cases are also shown, corresponding to more extreme combinations of $\mu$ and $\mathrm{Wa}$. The dead zones in current density are highlighted in the planar electrode in low temperature.}
    \label{fig:resistance_temp_fullcell}
\end{figure*}
We now focus on analysis of an interdigitated full cell, for fixed $H_\mathrm{ID}$ and $ H_\mathrm{gap}$, cf. \figref{fi:schematic}(B). 
The lengths of the interdigitated fin $L_{\mathrm{ID}}$ and the unshaped part of the electrode $L_\mathrm{bulk}$ are varied, such that the total volume of the electrode remains the same for all cases. 
Changing the length of the interdigitated fin $L_\mathrm{ID}$ is somewhat analagous to changing the amplitude of the interface in our previous half cell analysis, except for an important difference: the two electrodes are now allowed to become fully interwoven with one another. 
The interdigitated full cell design was automatically observed through the optimization work of \citet{li2024topology}, and here, we provide a physical explanation for the benefits of this design.

In  \figref{fig:resistance_temp_fullcell}(A), we show the relative resistance of the full cell as a function of $L_{\mathrm{ID}}$ for the $(\mu,\mathrm{Wa})$ values corresponding to low and room temperature operating conditions. 
Consistently with the half-cell analysis, the benefit of increasing the length of the features here is magnified at low temperature due to the current density accumulation at the electrode/electrolyte interface. 
Additionally, a greater benefit is seen with lower porosity electrodes. 
The interdigitated fins have a similar effect to the sinusoidal features previously considered, since they both bring the two electrodes effectively closer together. However, the reduction in resistance is larger for the interdigitated full cell, particularly when the fins interweave with one another, cf. the region shaded in grey. 
As the fins interweave, i.e. when they  cross the $x=0$ plane, the resistance reduction with respect to increasing $L_{\mathrm{ID}}$ slows. 
This is consistent with the length scale characterizing the electrode separation that transitions from $L_\mathrm{sep}$, the gap between the planar electrodes in the $x$ direction, to $(H_\mathrm{gap}-H_\mathrm{ID})/2$, the gap between the fins in the $y$ direction; after the electrodes  interweave, the gap between the fins in the $y$ direction no longer decreases with increasing $L_{\mathrm{ID}}$. 

To expand on this length scale discussion for the low temperature case, let us build intuition by considering the cell overpotential for planar electrodes. When $L_\mathrm{ID}=0$ and $\mu\rightarrow\infty$, the cell overpotential can be determined by solving  \eqnref{eq:phi1_nondim}-\eqnref{eq:phi2_nondim} in 1D and taking the limit. In particular,
\begin{equation} 
    \lim_{\mu\rightarrow\infty}\dfrac{\eta_\mathrm{cell}}{\mathcal{I}} = L_\mathrm{sep} + 2\dfrac{ \coth \left(\sqrt{\rho \mathcal{C} \mathrm{Wa}^{-1}} L_\mathrm{bulk}\epsilon ^{-3/4}\right)}{\sqrt{\rho \mathcal{C} \mathrm{Wa}^{-1} } \epsilon ^{3/4}},\label{eq:analytical_eta}
\end{equation}
where the first term on the right-hand side is the resistance through the electrolyte of length $L_\mathrm{sep}$ and the remaining term is the resistance through the two electrodes. 
This is consistent with the theoretical expression for linear polarization given by \citet{newman2021electrochemical}. 
As described before, the dominant length scale transitions from the $x$ direction to the $y$ direction as the  fins  interweave. In particular, the separation length scale  transitions from $L_\mathrm{sep}\rightarrow(H_\mathrm{gap}-H_\mathrm{ID})/2$, while the bulk electrode thickness length scale transitions from $L_\mathrm{bulk}\rightarrow H_\mathrm{ID}$. 
If we substitute these length scales into \eqnref{eq:analytical_eta},  the 
overpotential ratio is roughly equal to $0.31$; this $69\%$ reduction is mainly due to the decrease in electrolyte length scale. 
From  \figref{fig:resistance_temp_fullcell}(A), we see that the computed relative resistance in the low temperature case when $L_\mathrm{ID}=3$ is approximately $0.14$ ($86\%$ reduction): a reduction of similar magnitude.  Of course, this order-of-magnitude analysis is not meant to be quantitative, since the interdigitated geometry is more complex and $\mu$ is large, but certainly not $=\infty$. Nonetheless, this analysis highlights how the dominant length scales shift.

To gain further intuition on the benefits of interdigitation, we explore the root-mean-squared-deviation (RMSD) of the local current density $i_n$, i.e., a metric of current density uniformity in the electrode. Specifically, we define,
\begin{equation}
\mathrm{RMSD}(i_n) = \sqrt{\left\langle\left(\frac{i_n}{\langle i_n\rangle}-1\right)^2\right\rangle},
\end{equation}
where $\langle \cdot \rangle$ is the average over the porous electrode. 
In  \figref{fig:resistance_temp_fullcell}(B), we plot $\mathrm{RMSD}(i_n)$ as a function of $L_{\mathrm{ID}}$. 
We find that $\mathrm{RMSD}(i_n)$ generally decreases with increasing interdigitation, particularly for the low temperature case. 
This supports our intuition: interdigitation allows a more uniform distribution of current density and hence better utilization of the entire electrode.
\figref{fig:resistance_temp_fullcell}(D) shows the local normalized current distribution $i_n/\langle i_n\rangle$ for the low and room temperature planar and fully interdigitated electrode cases. To further justify our conjecture, current distributions corresponding to more extreme ($\mu$, $\mathrm{Wa}$) values are also shown. 
These values may not be obtainable by changing temperature with the given materials, but they may be obtained by using different materials.

In the \qty{-30}{\celsius} case, the current density in the planar electrode is localized to the electrode/electrolyte interface, leading to underutilized dead zones in the electrode interior. 
The current density is more evenly distributed for the interdigitated electrodes, consistent with the reduction of $\mathrm{RMSD}(i_n)$ from  \figref{fig:resistance_temp_fullcell}(B), thus explaining its dramatic reduction in resistance when compared to the planar electrode. 
For the room temperature case, the current density for the planar electrode is still relatively localized to the electrode/electrolyte interface. In principle, if we use materials with smaller $\mu$ and larger $\mathrm{Wa}$, such as the case on the right of  \figref{fig:resistance_temp_fullcell}(D), we further diminish the benefit to electrode shaping and increase the potential benefit to current collector shaping. Finally, observe that the current density distribution is qualitatively different for the planar and interdigitated cases as $(\mu,\mathrm{Wa})$ transitions from $(1,250)$ to $(1000,.25)$, i.e. to the left of  \figref{fig:resistance_temp_fullcell}(D). 
There is a transition from having pronounced gradients in the $x$ to the $y$ direction, supporting our prior discussion on length scales.

\section{Chemo-mechanics of shaped electrodes}\label{sec:mech}

In the previous section, we showed the importance of electrode shaping to overcome kinetic limitations in ion transport in porous electrodes. 
Here, we  analyze the mechanical response of the structures studied in \sectref{sec:electrostatics} to chemically, or similarly, thermally, induced expansion.
While the study presented in \sectref{sec:electrostatics} applies to both liquid and solid electrolyte systems, this chemo-mechanical analysis focuses on all-solid state batteries (ASSBs), where the electrolyte is a solid ion-conducting material.
Mechanical degradation is critical in ASSBs for two main reasons: (1) Ion transport and charge-transfer reactions can occur only if mechanical integrity is maintained in the bulk and at the interfaces; and (2) In a fully solid, low porosity cell, large mechanical stresses can develop upon cycling, because the chemical expansion of active materials is constrained by the electrolyte.  
During operation of an ASSB, the rigidity of the solid-solid interface causes mechanical stresses in response to the electrochemical reactions that take place.

The dynamic coupling between electrochemistry and mechanics in ASSBs has been the subject of several studies~\cite{bucci2017modeling,  koerver2018chemo, bucci2018mechanical, wang2021transitioning, kalnaus2023solid, bucci2017effect}. 
The combinations of material properties characterizing energy materials and specific battery chemistries is vast and beyond our scope. 
We aim to derive general guidelines relating shape and mechanical failure, and to do so, we make a few simplifications.  
We assume linear elasticity of an isotropic media and only consider the limiting conditions at the end of charge/discharge. 
Chemical expansion is modeled by an isotropic anelastic strain, characterized by a scalar chemical-expansion coefficient (also called the Vegard parameter) and a parameter describing the range of chemical compositions experienced by the active materials during cycling.

\subsection{Chemo-mechanics model} \label{sec:ECM_setup}
Under our above stated assumptions, the governing field equations that ensure mechanical equilibrium of our cell are
\begin{align}
    &\nabla \cdot \boldsymbol{\sigma}= 0, \label{eq:limMomentBalance} \\
    &\boldsymbol{\sigma} = \mathbf{C} : \left(\boldsymbol{\epsilon} - \boldsymbol{\epsilon}^{\text{an}}\right),
\end{align}
where $ \boldsymbol{\sigma}$ is the symmetric Cauchy stress tensor. The displacement $\mathbf{u}$ of the cell is confined at the locations indicated on \figref{fi:schematic}, i.e.
\begin{equation}
\mathbf{u}=0,
\end{equation}
and unless otherwise noted, the no-traction boundary condition, $ \boldsymbol{\sigma}\cdot \mathbf{n}=0$, is applied elsewhere. 
The anelastic strain and elasticity tensors are
\begin{align}
    & \boldsymbol{\epsilon}^{\text{an}} = \beta \left(c^\text{stoich} - c_0^\text{stoich} \right) \mathbf{I}, \\
    & \mathbf{C} = 2 G \mathbf{I}^\text{sym} + \lambda \mathbf{I}\otimes \mathbf{I},
\end{align}
where $\mathbf{I}$ is the Kronecker delta function, $\mathbf{I}^\text{sym}$ is the symmetric fourth order unit tensor\footnote{In index notation, $\mathbf{I}^\text{sym}$ is $\dfrac{1}{2} \left( \delta_{il} \delta_{jk} + \delta_{ik} \delta_{jl} \right)$
}, $c^\text{stoich} - c_0^\text{stoich}$ is the change in stoichiometry of the electrode material, $\beta$ is the isotropic Vegard coefficient, and $\lambda$ and $G$ are the Lam\'e parameters such that $\lambda=\nu E/[(1+\nu)(1-2\nu)]$ and  $G=E/[2(1+\nu)]$, where $E$ is the Young's modulus and $\nu$ is the Poisson's ratio.

The domain and boundary conditions for the chemo-mechanics problem are illustrated in \figref{fi:schematic}.
Given that the cell dimension in the $z$ direction is typically much larger than the thicknesses in the $x$ and $y$ directions, we assume plane-strain conditions. In other words, the cell is constrained from deforming out of plane, in the $z$ direction, i.e. $u_z=0$ and $\partial u_x/\partial z=\partial u_y/\partial z = 0$. 
It follows that $\sigma_{zz} = \lambda (\epsilon^\text{el}_{xx} + \epsilon^\text{el}_{yy})$, where $\boldsymbol{\epsilon}^\text{el}=\boldsymbol{\epsilon} - \boldsymbol{\epsilon}^{\text{an}}$. 
Similar to \sectref{sec:electrostatics}, \eqnref{eq:limMomentBalance} was solved in Wolfram Mathematica using a finite element method with \texttt{NDSolveValue} and \texttt{SolidMechanicsPDEComponent}.

We assume steady-state conditions, and a uniform composition: the electrode ranges from fully charged (corresponding to $c_0^\text{stoich} = 1$) to fully discharged ($c^\text{stoich}=0$).
The Vegard parameter $\beta$ describes the volumetric expansion associated with changes in chemical composition (analogous to thermal expansion).
For simplicity, we assume an isotropic Vegard's law, where
\begin{equation} \label{eq:vegard}
\beta = \begin{cases}
-\beta_1 &\text{in cathode},\\
\beta_1 &\text{in anode},\\
0 &\text{in solid electrolyte}.
\end{cases}
\end{equation}
Based on typical values for intercalating compounds, we assume a 3\% volume expansion of the composite electrode region, corresponding to a Vegard's parameter $\beta_1 = 0.01$ \cite{qi2014lithium}. 
The Young's modulus is 
\begin{equation} \label{eq:elasticP}
E = \begin{cases}
E_1 &\text{in cathode/anode},\\
E_2 &\text{in solid electrolyte},
\end{cases}
\end{equation}
where, unless otherwise noted, we choose $E_1=\qty{75}{GPa}$ and $E_2=\qty{25}{GPa}$~\cite{sakuda2013sulfide}. 
We assume the Poisson's ratio $\nu=0$ for both materials, so we can compare our results to the analytical solutions derived by \citet{hsueh1985residual} for residual stress in metal/ceramic bonded strips.

The assumption of homogeneous chemical expansion is significant, so we discuss its consequences. In general, this assumption is valid at low charging rates and for thin electrodes. 
Of course, a concentration gradient is generally expected and this gradient would affect the equilibrium in \eqnref{eq:limMomentBalance}.
As we have shown in \sectref{sec:electrostatics}, however, electrode shaping and interdigitation promote current uniformity which likely leads to lower strain gradients.
Additionally, if the mechanical response remains in the elastic regime, then the stress state is not dependent on the cycling history. 
Therefore, our steady-state problem reasonably captures the effects that the electrode shape, cell design, and boundary conditions have on the chemo-mechanic response of solid-state batteries from full charge to full discharge.

Finally, unlike our analysis in \sectref{sec:electrostatics}, where we found that varying the operating temperature within the range of $-30$ to \qty{50}{\celsius} has a significant effect on the electrochemical reactions, in this section, we ignore the effect of temperature: the mechanical response of battery materials is not particularly sensitive to temperature changes.
While not considered here, thermal strains arising during operation and material processing could in principle be added to the chemical strains. This can be significant in, e.g., solid-oxide cells that are operated at high temperatures ($>$ \qty{500}{\celsius}) \cite{ormerod2003solid}.

\subsection{Results and discussion} \label{sec:ECMresults}

\subsubsection*{Half-cell analysis}

\begin{figure}
    \centering
    \includegraphics[width=3.in]{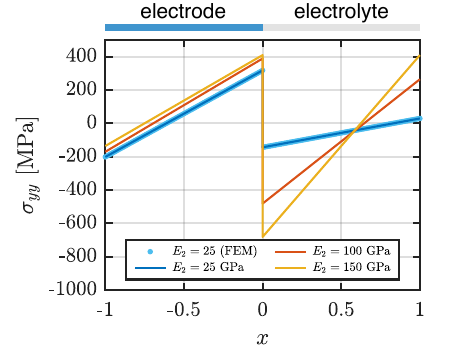}
    \caption{Comparison of the stress profile obtained with the finite element analysis (FEM, dotted line) and the analytical expression of residual stress in bi-layer materials under thermal strains~\cite{hsueh1985residual}. The analytical solution is employed to show the effect of electrolyte stiffness on the stress gradient. All the results are for a 2D half cell with a planar interface.}
    \label{fig:stress_vsE}
\end{figure}

\begin{figure*}[ht!]
    \centering
    \includegraphics[width=5.25in]{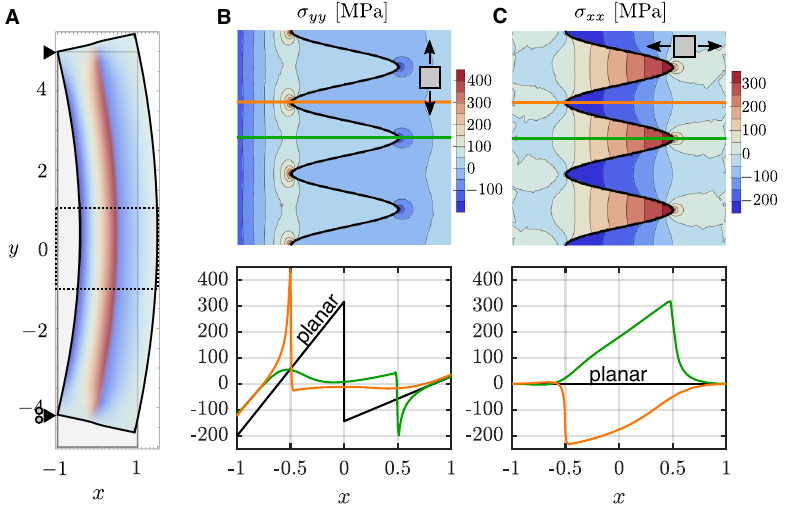}
    \caption{Stress fields for a half cell consisting  of an electrode undergoing chemically-induced volume change are shown. (A) Deformation and bending of the planar half cell caused by the negative volume change of the electrode is shown. (B, C) The $\sigma_{yy}$ and $\sigma_{xx}$ stress components are shown for a shaped electrode, respectively, when  the amplitude $A=0.5$ and  the frequency $f=3$. The bottom part of the figure shows the stresses as a function of $x$ along the lines shown in the contour plots above, i.e., at $y=1/f$ (orange) and $y=0$ (green). As a reference, the solution for the planar electrode case is also shown. In panel (B), the $\sigma_{yy}$ stress component is discontinuous at the interface, due to the difference in elastic properties (\eqnref{eq:elasticP}) and Vegard's parameter (\eqnref{eq:vegard}) in the two materials. In panel (C), with respect to $\sigma_{xx}$, we see that the electrode fins are under tension and the electrolyte fins under compression. For comparison, the $\sigma_{xx}$ stress is zero for the planar interface.} 
    \label{fig:HCchemoMech}
\end{figure*}

We analyze the effect of interface shape defined according to \eqnref{eq:shape_halfCell} on the stress field generated by chemical expansion/contraction of the composite electrode in a half cell. 
As an illustrative example, we assume that the electrode region undergoes a negative volume change. This occurs during ion de-intercalation at the cathode. 
The stress field follows the periodicity of the interface  and  does not change with frequency. Therefore, the results discussed here consider the effect of amplitude $A$, while keeping  the frequency $f$ constant. The displacement boundary conditions illustrated in \figref{fi:schematic} are representative of a simply-supported bi-layer structure. 

To verify our results, we first analyze planar ($A=0$) interfaces and compare to the analytical solution derived for  bi-layer slabs upon cooling~\cite{hsueh1985residual}.
\figref{fig:stress_vsE} shows that the $\sigma_{yy}$ stress profile computed from the finite-element analysis (dotted line) agrees with the analytical solution (solid blue line). 
In addition to verification, this analytical solution of~\citet{hsueh1985residual} can be used to demonstrate the stress distribution for stiffer electrolytes, showing that the stress gradient increases with the Young's modulus. 
The values of Young's modulus $E_2=\qty{25}{GPa}$~\cite{sakuda2013sulfide} and $E_2=\qty{150}{GPa}$~\cite{ni2012room} are representative of two classes of solid electrolyte materials, sulfide and oxide materials, respectively, characterized by almost one order of magnitude difference in stiffness.
The region in contact with the electrode is under compression, in all three cases shown in \figref{fig:stress_vsE}. However, the outer boundary ($x=1$) is nearly stress-free for the $E_2=\qty{25}{GPa}$ case, and undergoes an increasing tensile stress as  electrolyte stiffness increases. This difference in the stress gradient and its sign has significant effects on the mechanical, and consequently electrochemical, degradation of the cell.
The stiffer electrolyte better resists bending, which increases stress in the electrode.
It is important to note that the results in \figref{fig:stress_vsE} assume an elastic response. Analytical solutions for elasto-plastic materials were also derived by \citet{hsueh1985residual}, which could be used to explore plastic response, as a beneficial stress-release mechanism. 
However, plasticity analyses require the time history of the mechanical response throughout the battery cycling conditions. Such analyses are outside the scope of the present work.

We now compare the planar ($A=0$) and shaped ($A>0$) half cells. As shown in \figref{fig:HCchemoMech}(A), the electrode contraction is opposed by the electrolyte stiffness, causing the electrode to be under tension and the electrolyte to be under compression, resulting in bending of the half cell. 
\figref{fig:HCchemoMech}(B) and \figref{fig:HCchemoMech}(C) show the stress fields for the $A=0.5$ shaped cell.
Although $\sigma_{yy}$ is large in the tip regions, on average it is about one third of the baseline case of $A=0$.
The graph in \figref{fig:HCchemoMech}(B) shows the $\sigma_{yy}$ stress over the $y=0$ (green curve) and $y=1/f$ (orange curve) planes in comparison to the planar case (black curve). 
For the planar case, the $\sigma_{yy}$ stress is discontinuous at the interface, and within the electrode domain ($x<0$), $\sigma_{yy}$ changes linearly from tensile to compressive away from the interface, while it remains compressive within the electrolyte.
The peak $\sigma_{yy}$ stress values, of \qty{300}{MPa} in tension and \qty{200}{MPa} in compression, are approximately the same in the shaped and planar electrodes, but the high stress region is more localized in the shaped electrode.
The overall reduction in $\sigma_{yy}$ in the shaped electrode is compensated by a corresponding increase in $\sigma_{xx}$.
The contour plot in \figref{fig:HCchemoMech}(C) shows that the electrode protrusions are under tensile stress in the $x$ direction, i.e. $\sigma_{xx}>0$, as the material is contracting, while the electrolyte fins are under compressive stress, i.e. $\sigma_{xx}<0$. 
The difference in the stress between the electrode and electrolyte regions is apparent in the comparison of the two plots. The  $\sigma_{xx}$ stress is nearly anti-symmetric and increases almost linearly along the fins, reaching peak values at the fin tip. 
In contrast, $\sigma_{xx}$ is zero in the planar electrode.

\begin{figure}[t]
    \centering
    \includegraphics[width=3in]{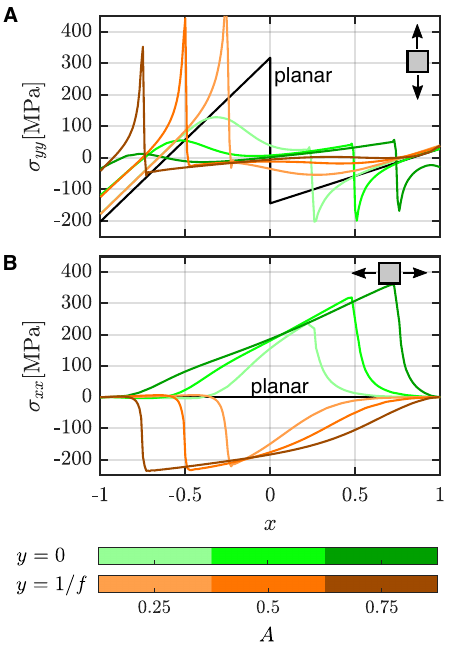}
    \caption{The $\sigma_{yy}$ and $\sigma_{xx}$ stresses are shown across the half-cell thickness, at different $y$ positions, and for varying shaping amplitude $A$, with $A\in\{0.25, 0.5, 0.75\}$. The reference case of a planar interface is marked in black. The green and orange curves mark the stress along the $y=0$ and $y=1/f$ planes, respectively.}
    \label{fig:stress_vsA}
\end{figure}

The sensitivity of the Cauchy stress to the shaping amplitude $A$ is explored in \figref{fig:stress_vsA}. The peak $\sigma_{yy}$ and  $\sigma_{xx}$ stress is in the tip region. 
The black curves show the planar interface stress for reference, and the green and orange curves show stresses along $y=0$ and $y=1/f$, respectively. 
The variation in green and orange tones represent the shaping amplitude, with darker colors indicating larger amplitudes. Qualitatively, the stress profiles do not change with $A$, however, the magnitude of the stress components change in favor of larger longitudinal stresses along the fins.
The peak $\sigma_{yy}$ tensile stress decreases with amplitude (from 490, to 440, to \qty{350}{MPa}), as the tips move closer to the boundary allowing for some stress release. 
However, the $\sigma_{xx}$ tensile  stress along the fins increases with increasing amplitude (from 240, to 320, to \qty{360}{MPa}). 
The tensile peaks remain located in the electrode regions, while the electrolyte domain is under compression.

These analyses reveal that shaped electrodes may introduce modes of mechanical failure by concentrating the stress in the high curvature regions, thus possibly nucleating fracture. 
If cracks grow in the direction parallel to the fins, i.e. the direction of the ion flux, they may not compromise the functionality of the cell. However, the specific microstructure of the composite electrode introduces a complexity that needs to be accounted for in order to obtain a more accurate determination of damage evolution. In the context of this chemo-mechanical analysis, we can conclude that shaped half-cell configurations do not present a mechanical benefit as compared to traditional planar 2D cells, as they promote stress concentration, which can lead to mechanical degradation and consequently to impedance growth. 

\begin{figure}
    \centering
    \includegraphics[width=3.25in]{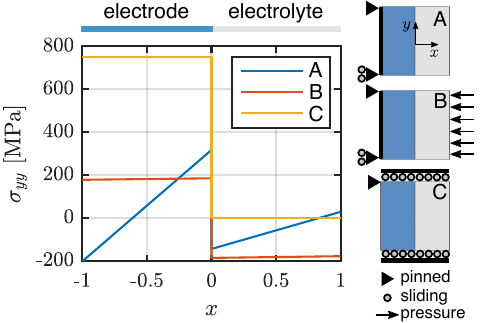}
    \caption{The $\sigma_{yy}$ stress is shown for a half cell comprised of an electrode ($x<0$) bonded to a solid-electrolyte layer ($x>0$) having the function of a separator. The electrode region undergoes a negative volume change caused by a change in chemical composition. The stress profile depends on the boundary conditions imposed on the half cell: case A represents a simply-supported beam consisting of two different materials bonded together; case B is similar to case A, but  includes a uniform stack pressure of \qty{3}{MPa}; and in case C, displacement in the $y$ direction is constrained.}
    \label{fig:chemMech_setup}
\end{figure}

The stress response also varies with the specific choice of applied boundary conditions (BCs). 
To illustrate this, \figref{fig:chemMech_setup} shows   $\sigma_{yy}$ stress for the planar cell corresponding to three different BCs, which are also illustrated schematically.
The analyses presented in \figref{fig:HCchemoMech} and \figref{fig:stress_vsA} correspond to case A, with the half cell allowed to bend out-of-plane while supported at the two ends;
note that the line representing case A is the same as it the appears in those figures.
Case B has the same displacement BCs as  case A, but a uniform stack pressure of \qty{3}{MPa} is applied orthogonal to the half-cell plane. 
In case C, the half cell is constrained from expanding/contracting in the $y$ direction.
In all cases, the $\sigma_{yy}$ stress is discontinuous across the interface, due to the difference in Young's modulus and anelastic strain, and the $\sigma_{xx}$ component is zero.
The application of the pressure load alters the $\sigma_{yy}$ stress profile, making the stress approximately equal in magnitude and opposite in sign in the electrode and electrolyte regions.
In case C, the electrode experiences a uniform tensile stress (caused by the negative volume change constrained at the edges), while the electrolyte region remains unloaded. This is due to the fact that the BCs prevent the cell from bending, and the stress in the electrode is independent of the solid-electrolyte stiffness.
In case A, instead, the electrolyte stiffness affects the bending curvature and the stress state in the electrode.
The analyses in \figref{fig:chemMech_setup} show that the stack pressure can lower the $\sigma_{yy}$ stress in the electrolyte. If the half cell, comprised of a cathode and a solid-state electrolyte layer, is combined with a Li-metal anode, the compression of the  electrolyte can mitigate opposing Li-protrusions and elecrodeposition-driven fracture.

\subsubsection*{Full-cell analysis}
We now extend our chemo-mechanics analysis to an interdigitated full cell, paralleling our analysis in \sectref{sec:electrostatics}. Again, we analyze the effect of varying the interdigitated fin length $L_\mathrm{ID}$, while changing the unshaped part of the electrode to keep the volumes of the electrode and electrolyte the same. Since changing $L_\mathrm{ID}$ is somewhat similar to varying the amplitude $A$ in the half-cell analysis, we focus on the effects of interpenetration. For this analysis, we consider the boundary conditions shown in  \figref{fi:schematic}(B), where the top and bottom faces are constrained in the $y$ direction, corresponding to case C in \figref{fig:chemMech_setup}.
For simplicity, we assume that both electrodes have the same mechanical properties and equal Vegard's parameter.

\begin{figure}[t]
    \centering
    \includegraphics[width=3.5in]{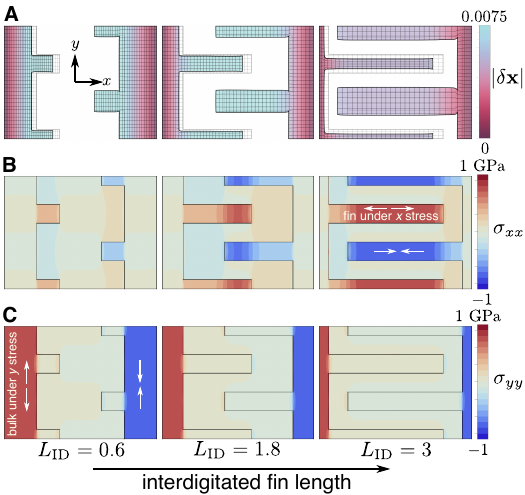}
    \caption{Chemo-mechanics analysis of interdigitated electrodes. (A) Local displacement $|\delta\mathbf{x}|$ is shown for the electrodes, where the original position of the electrodes is shown in grey. Displacements are exaggerated for the sake of visualization. The (B) normal stress in the $x$ direction, $\sigma_{xx}$, and (C) normal stress in the $y$ direction, $\sigma_{yy}$, are shown. The aspect ratio of the images are not representative of the actual aspect ratio.}
    \label{fig:stress_full}
\end{figure}

In  \figref{fig:stress_full}, we show solutions of mechanical equilibrium for various $L_\mathrm{ID}$, where the electrode on the left contracts due to ion de-intercalation and the electrode on the right expands due to intercalation. When $L_\mathrm{ID}=0$, i.e. planar electrodes, the solution is effectively one-dimensional, with the electrolyte region simply translating toward the contracting electrode. As $L_\mathrm{ID}$ increases, the deformation becomes more complex, particularly near the corners of the interdigitated fins, as shown in  \figref{fig:stress_full}(A). The $x$ and $y$ components of the normal stress, $\sigma_{xx}$ and $\sigma_{yy}$, are shown in \figref{fig:stress_full}(B, C), respectively. From this analysis, we observe that the contracting interdigitated fins are under tension in the $x$ direction, while the expanding fins are under compression in the $x$ direction, with the magnitude of these stresses increasing with $L_\mathrm{ID}$. Physically, this is a result of the electrolyte acting as an elastic medium resisting the contraction and expansion. Additionally, the bulk part of the contracting electrode is under tension in the $y$ direction, while the bulk part of the expanding electrode is under compression. This is due to a similar physical reason, except the resistance to contraction and expansion is due to the solid walls, i.e. the boundary conditions on the top and bottom faces; this is consistent with the previous discussion of the BCs in case C in \figref{fig:chemMech_setup}.

\begin{figure}[t]
    \centering
    \includegraphics[width=3.5in]{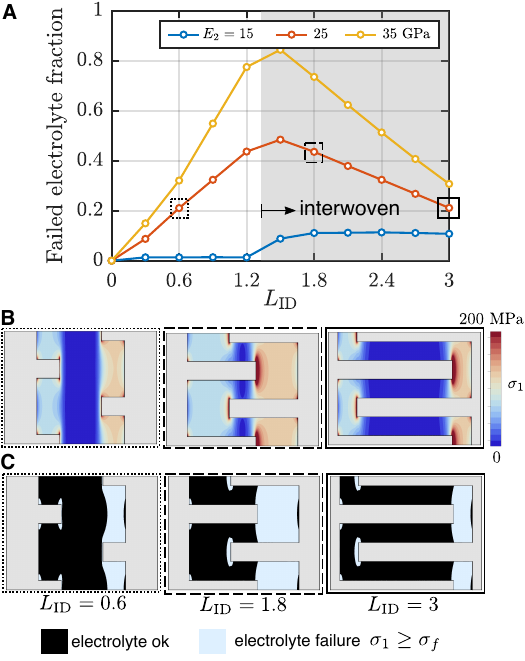}
    \caption{(A) The fraction of solid electrolyte that exceeds the failure criterion in \eqnref{eq:fail} are plotted as a function of interdigitated fin length $L_\mathrm{ID}$ for various Young's moduli of the electrolyte $E_2$. Corresponding images are shown depicting the (B) principal stress $\sigma_1$ and (C) locations where the electrolyte fails, for the case where the electrolyte Young's modulus is $E_2=\qty{25}{GPa}$. The aspect ratio of the images are not representative of the actual aspect ratio.}
    \label{fig:elyte_fail}
\end{figure}

Next, we analyze how failure occurs in the solid electrolyte as a function of $L_\mathrm{ID}$. When the electrodes are planar, this is not a concern, since the electrolyte simply translates. From the stress tensor, we calculate the principal stresses
\begin{equation}
\sigma_{1,2} = \frac{\sigma_{xx}+\sigma_{yy}}{2} \pm \sqrt{\left(\frac{\sigma_{xx}-\sigma_{yy}}{2}\right)^2 + \sigma_{xy}^2},
\end{equation}
i.e., the stress eigenvalues, throughout the electrolyte. Assuming that the solid electrolyte is a brittle material that primarily fails in tension, we denote our electrolyte failure criteria as
\begin{equation}
\sigma_1\geq \sigma_f, \label{eq:fail}
\end{equation}
where $\sigma_f$ is a fracture strength, here assumed to be $\sigma_f=\qty{100}{MPa}$. In \figref{fig:elyte_fail}(A), we show the fraction of the electrolyte that exceeds the failure criteria in \eqnref{eq:fail} as a function of $L_\mathrm{ID}$, for a variety of electrolyte Young's moduli $E_2$, where the region shaded in grey indicates the lengths of  $L_\mathrm{ID}$ corresponding to interwoven electrodes.
For the case where the electrolyte Young's modulus is $E_2=\qty{25}{GPa}$, in \figref{fig:elyte_fail}(B), the corresponding images show the fields of principal stress $\sigma_1$, and in \figref{fig:elyte_fail}(C), the specific locations within the electrolyte where failure occurs are highlighted in light blue.
As expected, stiffer electrolyte leads to a larger failure fraction.  Interestingly, we observe that the fraction of failed electrolyte is not monotonic in $L_\mathrm{ID}$. When $L_\mathrm{ID}=0$, none of the electrolyte fails, since the electrolyte simply translates as a block. As $L_\mathrm{ID}$ increases, the fraction of the failure increases, primarily localized in the gap between the fins of the expanding electrode. However, as $L_\mathrm{ID}$ increases further, the electrodes become interwoven, qualitatively changing the stress pattern. The stress within the gap between the interdigitated fins reduces below the failure criteria: since one fin is expanding near another contracting fin, the electrolyte largely translates in this region instead of being significantly strained. 

Overall, while planar electrodes may be the least likely to lead to electrolyte damage, this analysis provides evidence that the geometry that would be most beneficial from the electrical cell resistance perspective -- fully interdigitated electrodes -- also avoids the worst of the electrolyte stresses. Similarly, this analysis suggests that fully interwoven electrodes may perform better mechanically than two individually shaped electrodes that are separated by a planar separator. 
However, there is certainly still a trade-off between resistance and mechanics, and depending on the requirements of a specific application, these pros and cons should be balanced accordingly.

\section{Concluding remarks}\label{sec:conc}
In this work, we explored the effects of shape in porous electrodes in order to elucidate the associated trade-offs between electrical and mechanical performance. Inspired by the variety of cell architectures that have been previously studied, we examined both a sinusoidal half cell and an interdigitated full cell. The sinusoidal shape introduces low-tortuosity channels of pure electrolyte that facilitate ion transport. The interdigitated full cell leverages this same idea, but also changes the separation length-scale between the electrodes.

First, we used a simple electrostatics model to analyze cell resistances as a function of shape. We focused our attention on low temperature performance, where the electrolyte conductivity decreases, and ion transport in the electrolyte becomes the limiting kinetics. 
Our model describes the cell kinetics primarily through two dimensionless groups, the ratio of electric to ionic conductivity $\mu$, and the Wagner number $\mathrm{Wa}$, i.e., the ratio between surface reaction and Ohmic resistance. The low-temperature regime is characterized by larger $\mu$ and lower $\mathrm{Wa}$ values. 
Overall, we find the benefits of electrode shaping to be amplified in this regime, as the reaction becomes primarily localized at the electrolyte-porous electrode interface. 
While electrode structures tend to reduce cell resistance, the current distribution shows that interdigitation mitigates the presence of reaction dead zones at low temperature. 

Next, we leveraged a simple chemo-mechanics model to study how the same electrode shapes respond to the mechanical loading caused by chemical contraction/expansion during battery cycling. While the electrostatics analysis is applicable to both liquid and solid-state batteries, for this section we focus on the latter, given that mechanical degradation can be more severe and consequential with solid-solid interfaces.
We found that shaped electrodes do not alleviate stress, as compared to traditional planar designs. Instead, they promote stress localization within the electrode fins and at the tips. This may result in fracture nucleation and further studies focused on damage are required.
In an interdigitated full cell, we find that the stress in the fins is aligned with the direction of the fins themselves, whereas the bulk electrode is under stress in the transverse direction. By analyzing the principal stresses in the electrolyte, we find that the maximum fraction of failed electrolyte occurs when the interdigitated fins are about to become interwoven, with the fraction decreasing with increasing fin length. 
In summary, fully interdigitated electrodes provide the most benefit in terms of transport kinetics and tolerable stress. 

Our analysis has leveraged simple physics models to demonstrate the trade-off between electrical and mechanical performance when shaping electrodes. To do so, several assumptions have been made that could be addressed as future work. For example, we have neglected mass transfer and time-dependence: it would be of interest to understand how our conclusions change as the concentration fields of the reactant are allowed to deplete. As a stepping stone to understanding the effect of mass transfer, a spatially dependent concentration field could be imposed as $\mathcal{C}=\mathcal{C}(\mathbf{x})$ in  \eqnref{eq:dimgroups2}, instead of just assuming $\mathcal{C}=1$. Furthermore, in our analysis of the electrostatics of the interdigitated full cell, we have assumed that the anode and cathode are symmetric in their material properties. Of course, this is not required, and allowing for asymmetric material properties may show that the ideal electrode shape may be different for anode and cathode. Finally, in our chemo-mechanics analysis, we have assumed a uniform expansion and contraction; of course, this is not entirely realistic, since redox reaction and ion intercalation does not need to occur uniformly. If more expansion and contraction occurs near the electrode-electrolyte interface, the stress patterns would be different. Understanding effects such as these, and particularly how they affect performance trade-offs, will be key in improving shaped battery electrode design.

\section*{Acknowledgments}
This work was performed under the auspices of the U.S. Department of Energy by the Lawrence Livermore National Laboratory under contract DE-AC52-07NA27344. This work was supported by the Lawrence Livermore National Laboratory LDRD 23-SI-002. LLNL release number: LLNL-JRNL-864862.

\section*{List of main symbols}

\subsection*{Dimensional}
\begin{description}
    \item[$a$] specific surface area [m$^{-1}$]
    \item[$\mathbf{C}$] elasticity tensor [Pa]
    \item[$E_1$] Electrode Young's modulus [Pa]
    \item[$E_2$] Electrolyte Young's modulus [Pa]
    \item[$F$] Faraday's constant [C/mol]
    \item[$i_n$] current density [mA/cm$^2$]
    \item[$i_0$] exchange current density [mA/cm$^2$]
    \item[$\kappa$] electrolyte conductivity [mS/cm]
    \item[$L$] half-cell thickness [m]
    \item[$R$] gas constant [J/K mol]
    \item[$\sigma$] electrode conductivity [mS/cm]
    \item[$\boldsymbol{\sigma}$] stress tensor [Pa]
    \item[$T$] temperature [K]
    \item[$U_0$] standard potential [V]
    \item[$\mathbf{u}$] displacement [m]
\end{description}

\subsection*{Dimensionless}
\begin{description}
    \item[$A$] sinusoidal interface amplitude
    \item[$\alpha$] transfer coefficient
    \item[$\beta$] Vegard coefficient
    \item[$c^\mathrm{stoich}$] electrode stoichiometry 
    \item[$\mathcal{C}$] concentration
    \item[$\epsilon$] porosity
    \item[$\boldsymbol{\epsilon}$] strain tensor
    \item[$\eta$] overpotential
    \item[$f$] sinusoidal interface frequency
    \item[$\mathcal{I}$] applied current density
    \item[$L_\mathrm{ID}$] interdigitated fin length
    \item[$\mu$] solid/liquid conductivity ratio
    \item[$\phi_1$] solid-phase potential
    \item[$\phi_2$] liquid-phase potential
    \item[$\rho$] roughness
    \item[$\mathrm{Wa}$] Wagner number
\end{description}

\bibliographystyle{unsrtnat}
\bibliography{electstruct.bib}

\end{document}